\documentclass{epl1}
\usepackage{amsmath}
\usepackage{amssymb}
\usepackage{epsfig,float,afterpage,wrapfig,psfrag}

\newcommand{\be}{\begin{equation}}
\newcommand{\cc}{\mbox{\scriptsize{c}}}
\newcommand{\ii}{\mbox{\scriptsize{i}}} 

\newcommand{\ee}{\end{equation}}
\newcommand{\bea}{\begin{eqnarray}}

\newcommand{\eea}{\end{eqnarray}}

\newcommand{\ssz}{\scriptsize}

\newcommand{\scrF}{{\cal F}}

\newcommand{\w}{\omega}
\newcommand{\K}{\mbox{\scriptsize{K}}}
\newcommand{\m}{\mbox{\scriptsize{m}}}
\newcommand{\wm}{\w_{\m}}
\newcommand{\im}{\mbox{i}}
\newcommand{\subi}{\mbox{\scriptsize{I}}}
\newcommand{\subr}{\mbox{\scriptsize{R}}}

\DeclareMathAlphabet{\bi}{OML}{cmm}{b}{it}
\newcommand{\kk}{\bi{k}}

\renewcommand{\theequation}{\arabic{equation}}

\newcounter{saveeqn}

\title{Spectral scaling and quantum critical behaviour in the pseudogap
 Anderson model.}
\author{M. T. Glossop and D. E. Logan}
\institute{Oxford University, Physical and Theoretical Chemistry Laboratory,
South Parks Road, Oxford OX1 3QZ, UK}
\shorttitle{Dynamics of the pseudogap Anderson model}
\begin{document}
\maketitle
\begin{abstract}
The pseudogap Anderson impurity model provides a classic example of an
essentially local quantum phase transition. Here we study its single-particle 
dynamics in the vicinity of the symmetric quantum critical point
(QCP) separating generalized Fermi liquid and local moment phases, via
the local moment approach.  Both phases are shown to be characterized by
a low-energy scale that vanishes at the QCP; and the universal scaling spectra,
on \emph{all} energy scales, are obtained analytically. The 
spectrum precisely at the QCP is also obtained; its form showing clearly
the non-Fermi liquid, interacting nature of the fixed point.
\end{abstract}
The celebrated Anderson impurity model (AIM) \cite{b.1}, reviewed in \cite{b.2},
is the paradigm of quantum impurity physics. It describes a non-degenerate
impurity with energy $\epsilon_{\ii}$ and local interaction $U$, hybridizing
via $V_{\ii\kk} \equiv V$ to a non-interacting host with density of
states $\rho(\omega) = \sum_{\kk} \delta(\omega - \epsilon_{\kk})$
(and $\omega =0$ the Fermi level). The Hamiltonian in standard notation is

\begin{equation}
\hat{H} = \sum_{{\kk},\sigma}\epsilon_{\kk}\hat{n}_{{\kk} \sigma} +
\sum_{\sigma}(\epsilon_{\ii} + \tfrac{1}{2}U\hat{n}_{\ii-\sigma})\hat{n}_{\ii\sigma} +
\sum_{{\kk},\sigma}V_{\ii{\kk}}(c_{\ii\sigma}^{\dagger}c_{{\kk}\sigma} + h.c.).
\end{equation}
For a conventional metallic host, $\rho(0) \neq 0$, the problem is rather 
well understood \cite{b.2}, the impurity spin being quenched and the system ubiquitously
a Fermi liquid for all $U$. Its large-$U$ behaviour is that of the Kondo effect, 
characterised by the single low-energy scale $\omega_{\K}$; as embodied
famously in the Kondo resonance in the impurity single-particle spectrum
$D(\omega)$, which scales universally in terms of $\omega/\omega_{\K}$ alone.

  The underlying physics is much richer if the host contains a power-law
pseudogap \cite{b.3}, $\rho(\omega) \propto |\omega|^{r}$, $r>0$. This model has 
been studied extensively in recent years [3-12], especially for large-$U$ 
where the low-energy subspace maps onto that of the Kondo model. It
contains in general a non-trivial quantum phase transition at a 
finite $U_{\cc} \equiv U_{\cc}(r)$; the quantum
critical point (QCP) separating a degenerate local moment (LM) 
phase arising for $U>U_{\cc}$, from a `strong coupling' or generalized Fermi 
liquid (GFL) state [3-12] in which the impurity spin is locally quenched and 
a Kondo effect is  manifest. As $U \rightarrow U_{\cc}-$ the Kondo scale
$\omega_{\K}$ vanishes and the local spectrum again exhibits universal 
$\omega/\omega_{\K}$-scaling \cite{b.8,b.9}. The QCP itself, where the weight of the 
Kondo resonance has just vanished, has been accessed very recently \cite{b.12}
via a study of the \emph{local} magnetization and susceptibility. The fixed
point was found \cite{b.12} to be interacting, exhibiting critical local moment fluctuations and an associated destruction of the Kondo effect very similar to those present at the local QCP of the Kondo lattice \cite{b.13}, recently invoked to explain the behaviour of heavy fermion metals such as  
 CeCu$_{6-x}$Au$_{x}$ \cite{b.13}.  This provides topical motivation for further study of the model in the broad context of quantum phase transitions \cite{b.13a}, as too does its recently shown applicability to the issue of impurity moments in $d$-wave superconductors \cite{b.11}.

  Here we study single-particle dynamics of the pseudogap AIM [6,8--11] via the local 
moment approach (LMA) \cite{b.14}, a physically motivated many-body theory that
has already led \cite{b.8} to new predictions for the problem, borne
out by direct numerical renormalization group (NRG) calculations \cite{b.9}. Our
aims here, where we focus on the large-$U$ Kondo regime of
the symmetric model, are threefold. First to obtain the
universal scaling spectrum in the GFL phase: \emph{analytically} and
on  \emph{all} $\omega/\omega_{\K}$-scales, including but going far beyond 
the low-energy confines of traditional Fermi liquid theory. This is a
non-trivial issue even for the metallic AIM ($r=0$); where only recently
has the scaling spectrum for $\omega/\omega_{\K} \gtrsim 1$ been shown \cite{b.15}
to be dominated not by power-law decays as long thought; but instead
by long, slowly varying logarithmic tails. As we show,
the situation is much richer in the pseudogap case, and 
significantly different from previous expectations \cite{b.8,b.9}. Second,
in the LM phase where the Kondo scale $\omega_{\K} =0$, we consider
two questions: do the 
dynamics exhibit universal scaling close to the QCP 
on the LM side; if so what is the relevant low-energy scale, its physical origin and the form 
of the scaling spectra? Finally, we determine analytically  the behaviour of 
the single-particle spectrum precisely \emph{at} the QCP. Its \emph{low}-energy
behaviour is in turn shown to be intimately related to the \emph{high}-energy
asymptotics of the scaling spectra for both the GFL and LM phases; and 
its form shows clearly the non-Fermi liquid, interacting nature of the
fixed point.

\section{Local moment approach}  
We consider the symmetric QCP that arises in both particle-hole (p-h)
symmetric and asymmetric models \cite{b.5,b.7}. To that end we here focus explicitly on the 
p-h symmetric AIM with $\epsilon_{\ii} = -\frac{U}{2}$ (and local charge
$n_{\ii} = \sum_{\sigma}\langle\hat{n}_{\ii\sigma}\rangle =1$ $ \forall \  U$); 
where the QCP separates GFL/LM phases for all $0<r<\frac{1}{2}$ [5--9]. 
The host band is described by the
(simplified) form $\rho(\omega) = \rho_{0}|\omega|^{r}\theta(1-|\omega|)$
with bandwidth $D \equiv 1$ ($\theta(x)$: unit step function). We consider
the $T=0$ impurity Green function $G(\omega)$ and hence local spectrum
$D(\omega)= -\tfrac{1}{\pi}\mbox{sgn}(\omega)$Im$G(\omega)$; written conventionally as
$G(\omega) = [\omega^{+}-\Delta(\omega)-\Sigma(\omega)]^{-1}$, with
$\omega^{+}=\omega + \im 0^{+}\mbox{sgn}(\omega)$ and $\Sigma(\omega) =
\Sigma^{R}(\omega) -\im \mbox{sgn}(\omega)\Sigma^{\subi}(\omega)$ the single self-energy
(defined to exclude the trivial Hartree term).
$\Delta(\omega) = \Delta_{\subr}(\omega) -i\mbox{sgn}(\omega)\Delta_{\subi}(\omega)$ 
is the host-impurity hybridization, with $\Delta_{\subi}(\omega) = 
\pi V^{2}\rho(\omega) \equiv \Delta_{0}|\omega|^{r}\theta(1-|\omega|)$ where 
$\Delta_{0} = \pi V^{2}\rho_{0}$; and $\Delta_{\subr}(\omega)$ follows by
Hilbert transformation, $\Delta_{\subr}(\omega) = 
-\mbox{sgn}(\omega)[\beta(r)\Delta_{\subi}(\omega) + {\cal O}(|\omega|)]$ with
$\beta(r) = \mbox{tan}[\tfrac{\pi}{2}r]$.

  The most illuminating expos\'e of single-particle dynamics in the
GFL phase resides in the modified spectral function 
${\cal{F}}(\omega) = \pi\Delta_{0}\mbox{sec}^{2}(\tfrac{\pi}{2}r)
|\omega|^{r} D(\omega)$ [8--10]; such that at the Fermi level 
${\cal{F}}(\omega=0)=1$ \cite{b.10}. The latter result is exact, generalizing to $r>0$
behaviour well known for the metallic AIM $r=0$ (where it amounts essentially
trivially to satisfaction of the Friedel sum rule \cite{b.2}). It embodies the
fact \cite{b.8,b.10} that the leading low-$\omega$ behaviour, $\pi \Delta_{0}D(\omega)
\sim \mbox{cos}^{2}(\tfrac{\pi}{2}r)|\omega|^{-r}$, is precisely that of the 
non-interacting limit: $\Sigma^{\subr/\subi}(\omega)$ vanishes as 
$\omega \rightarrow 0$ more rapidly than the hybridization 
($\propto |\omega|^{r}$), reflecting the perturbative
continuity to the $U=0$ limit that in essence defines the GFL state.
The Kondo resonance is \emph{directly} apparent in $\cal{F}(\omega)$ [8--10]
(see also Fig.1 below). Determined by the Kondo scale $\omega_{\K}$, it 
narrows with increasing $U$ as the GFL/LM 
transition is approached and $\omega_{\K} \rightarrow 0$. And the LMA 
predicts \cite{b.8} $\cal{F}(\omega)$ itself to be universal in $\omega/\omega_{\K}$, 
as supported by NRG calculations \cite{b.9}. 

  The LMA for the pseudogap AIM is detailed in \cite{b.8}. Rather than calculating
$\Sigma(\omega)$ directly it employs a two-self-energy description: the 
rotationally invariant $G(\omega)$ is written formally as
$G(\omega) = \tfrac{1}{2}\sum_{\sigma}G_{\sigma}(\omega)$ with
$G_{\sigma}(\omega) = [\omega^{+} - \Delta(\omega) - 
\tilde{\Sigma}_{\sigma}(\omega)]^{-1}$. The self-energies 
$\tilde{\Sigma}_{\sigma}(\omega)$
($= -\tilde{\Sigma}_{-\sigma}(-\omega)$ by p-h symmetry)
are separated as $\tilde{\Sigma}_{\sigma}(\omega) =
-\tfrac{\sigma}{2}U|\mu| + \Sigma_{\sigma}(\omega)$, into
a static Fock piece with local moment $|\mu|$ that would alone
survive at pure mean-field (MF) level; together with the dynamical
$\Sigma_{\sigma}(\omega) = \Sigma_{\sigma}[\{\cal{G}_{\sigma}\}]$
that naturally dominate the low-energy physics and are functionals of 
the broken symmetry MF propagators
$\cal{G}_{\sigma}(\omega)$$ = [\omega^{+}-\Delta(\omega)
+\sigma\tfrac{U}{2}|\mu|]^{-1}$. In practice \cite{b.8,b.14} the LMA includes
in $\Sigma_{\sigma}(\omega)$ a non-perturbative class of diagrams
that embody coupling of single-particle excitations to low-energy
spin-flip processes, which is the essence of Kondo physics and
provides a mechanism for dynamical communication between the broken
symmetry MF states. These diagrams, shown \it e.g. \rm in Fig 3
of \cite{b.8}, translate for $\Sigma_{\uparrow}(\omega)$ to

\begin{equation}
\Sigma_{\uparrow}(\omega) = U^{2}\int^{\infty}_{-\infty} \frac{d\omega_{1}}
{\pi} \ \mbox{Im}\Pi^{+-}(\omega_{1})\left[\theta(\omega_{1}){\cal{G}}_{\downarrow}^{-}(
\omega_{1}+\omega) + \theta(-\omega_{1}){\cal{G}}_{\downarrow}^{+}(\omega_{1}
+ \omega)\right]
\end{equation} 
where ${\cal{G}}_{\sigma}^{\pm}(\omega)$ denote the one-sided Hilbert
transforms of 
the MF spectra \\ $D_{\sigma}^{0}(\omega)= -\tfrac{1}{\pi}\mbox{sgn}(\omega)$Im${\cal{G}}
_{\sigma}(\omega)$. $\Pi^{+-}(\omega)$ is the transverse spin polarization 
propagator, given at the simplest level employed here by an RPA-like
p-h ladder sum (see \it e.g. \rm Fig 3 of \cite{b.8}).

  In describing the GFL phase the final, central idea behind the LMA is
that of \emph{symmetry restoration} \cite{b.8,b.14,b.15}: self-consistent restoration of the
broken symmetry endemic at pure MF level, as embodied in 
$\tilde{\Sigma}_{\uparrow}(0)=\tilde{\Sigma}_{\downarrow}(0)$ and
hence (\it via \rm p-h symmetry)

\begin{equation}
\tilde{\Sigma}_{\uparrow}(0) = \Sigma_{\uparrow}(0) - \tfrac{1}{2}
U|\mu| = 0
\end{equation}
which ensures the Fermi liquid behaviour ${\cal{F}}(0)=1$. This is
achieved in practice \cite{b.8} for given $\tilde{U} = {U}/{\Delta_{0}}$
by varying the local moment $|\mu|$ from its pure MF value; which in turn
introduces a low-energy scale $\omega_{\m} \equiv \omega_{\m}(r)$ into the
problem through a strong resonance in Im$\Pi^{+-}(\omega)$ at
$\omega = \omega_{\m}$. This is the Kondo or spin-flip scale, 
$\omega_{\m} \equiv \omega_{\K}$; and it sets the timescale $\tau \sim
h/\omega_{\m}$ for symmetry restoration. In the LM phase by contrast, 
symmetry restoration Eq.(3) is not satisfied and $\omega_{\m}=0$.
Im$\Pi^{+-}(\omega)$ here contains a pole at $\omega =0$ reflecting the
zero energy spin-flip cost symptomatic of the locally degenerate phase; and the
`renormalized levels' $\tilde{\Sigma}_{\sigma}(0)$ are non-zero and
sign-definite \cite{b.8}. The GFL/LM phase boundary $\tilde{U}=\tilde{U}_{\cc}(r)$
may be obtained either from the limit of solutions to Eq.(3) (as
$\tilde{U} \rightarrow \tilde{U}_{\cc}-$); or, coming from the LM phase
and yielding the same $\tilde{U}_{\cc}$, from the condition that
$\tilde{\Sigma}_{\sigma}(0) \rightarrow 0$. GFL and LM phases are
correctly found \cite{b.8} to arise for all $0< r < \tfrac{1}{2}$, and solely LM 
states for $r> \frac{1}{2}$ and all $\tilde{U} >0$.

\section{GFL scaling spectrum} Close to the QCP, 
${\cal{F}}(\omega)$ scales in terms of $\tilde{\omega} = \omega/\omega_{\m}$
\cite{b.8,b.9}. To obtain the scaling behaviour one formally considers finite
$\tilde{\omega}$ in the limit $\omega_{\m} \rightarrow 0$; thus projecting
out non-universal features such as the Hubbard satellites. The hybridization
$\Delta(\omega) = \Delta(\tilde{\omega}\omega_{\m})$ then reduces to
$(\Delta_{0}\omega_{\m}^{r})^{-1}\Delta(\omega) =
-\mbox{sgn}(\omega)[\beta (r)+\im]|\tilde{\omega}|^{r}$, and the `bare'
$\omega = \omega_{\m}\tilde{\omega}$ in $G_{\sigma}(\omega) =
[\omega - \Delta(\omega) - \tilde{\Sigma}_{\sigma}(\omega)]^{-1}$ may
be neglected (only $r < \tfrac{1}{2}$ is relevant). Since
${\cal{F}}(\omega) = \pi \Delta_{0} \mbox{sec}^{2}(\tfrac{\pi}{2}r)
|\tilde{\omega}|^{r}\omega_{\m}^{r}D(\omega)$ exhibits scaling
so too does $\omega_{\m}^{r}D(\omega)$, given for $\omega \geq 0$ by

\begin{equation}
\Delta_{0}\omega_{\m}^{r}D(\omega) = -\mbox{$\frac{1}{2\pi}$Im}
\sum_{\sigma}[(\beta (r)+\im)|\tilde{\omega}|^{r} -
(\Delta_{0}\omega_{\m}^{r})^{-1}\tilde{\Sigma}_{\sigma}(\omega)]^{-1}.
\end{equation}
Eq.(4) also shows that $(\Delta_{0}\omega_{\m}^{r})^{-1}\tilde{\Sigma}
_{\sigma}(\omega)$ must scale universally in terms of $\tilde{\omega}$;
to which we now turn within the LMA, following the approach of \cite{b.15}
for the metallic AIM. Specifically we consider the strong
coupling Kondo regime where $1 \ll \tilde{U} $ ($< \tilde{U}_{\cc}(r)$ 
for the GFL phase). Since $\tilde{U}_{\cc}(r) \sim
\tfrac{1}{r}$ as $r \rightarrow 0$, our analysis holds 
strictly for small $r$, although the central conclusions remain 
qualitatively valid save for $r \rightarrow \tfrac{1}{2}$ (which
the approach below does not handle).

  In strong coupling where the local moment saturates ($|\mu| \rightarrow 1$),
the transverse spin polarization propagator entering Eq.(2) reduces to
$\tfrac{1}{\pi}$Im$\Pi^{+-}(\omega) = \delta (\omega -\omega_{\m})$. Hence
$\Sigma_{\uparrow}(\omega) = U^{2}{\cal{G}}_{\downarrow}^{-}(\omega + 
\omega_{\m})$, and in particular ($\tilde{\Sigma}_{\uparrow}^{\subi}(\omega) 
\equiv$) $\Sigma_{\uparrow}^{\subi}(\omega) = \theta (-[\omega + \omega_{\m}])
\pi U^{2}D_{\downarrow}^{0}(\omega + \omega_{\m})$. The leading low-$\omega$
dependence of the MF $D_{\downarrow}^{0}(\omega)$, which alone is relevant
to the scaling behaviour, is $D_{\downarrow}^{0}(\omega) \sim
({4\Delta_{0}}/{\pi U^{2}}) |\omega|^{r}$. Hence 

\begin{equation}
(\Delta_{0}\omega_{\m}^{r})^{-1} \tilde{\Sigma}_{\uparrow}^{\subi}(\omega) =
\theta (-[\omega + \omega_{\m}]) 4 |1+\tilde{\omega}|^{r}
\end{equation}
which indeed depends solely upon $\tilde{\omega}$ with no dependence on bare 
material parameters. The leading relevant low-$\omega$ behaviour of
$\mbox{Re}{\cal{G}}_{\downarrow}^{-}(\omega)$ is likewise given by \cite{b.8}

\begin{equation}
U^{2}\mbox{Re}{\cal{G}}_{\downarrow}^{-}(\omega) \sim \Delta_{0}\left[\frac{4}{\pi r} - \gamma (r) |\omega|^{r}\right]
\end{equation}
where $\gamma (r) = {4}/{\mbox{sin}[\pi r]} \sim {4}/{\pi r}$ (
and a uv-cutoff of order $D \equiv 1$ has been employed \cite{b.8}). 

The dependence
upon bare parameters of the Kondo scale $\omega _{\m}(r)$ and critical $U_{\cc}(r)$
follows from symmetry restoration Eq.(3), viz
($U^{2}$Re${\cal{G}}_{\downarrow}^{-}(\omega _{\m}) =$) $\Sigma _{\uparrow}^{\subr}
(0) = \tfrac{U}{2}$ in strong coupling. The critical $U_{\cc}$ where
$\omega _{\m}=0$ follows using Eq.(6) as $\tilde{U}^{-1}_{\cc}(r)
= \tfrac{\pi r}{8}$; which is exact as $r \rightarrow 0$ \cite{b.8,b.9}. (For the 
corresponding Kondo model, the exchange coupling $J$ relates to the
AIM parameters under a Schrieffer-Wolff transformation by
$\rho _{0}J = 8/\pi \tilde{U}$; whence the critical
$\rho _{0}J_{\cc} = 8/\pi \tilde{U}_{\cc} = r$, as obtained originally
by a large-$N$ study of the pseudogap Kondo model \cite{b.3}). For 
$\tilde{U} < \tilde{U}_{\cc}$, the $\tilde{U}$-dependence of $\omega_{\m}$ 
follows from symmetry
restoration as $\omega_{\m}(r) = 
(1-[\tilde{U}/\tilde{U}_{\cc}])^{1/r} =
(1-\tfrac{\pi r}{8}\tilde{U})^{1/r}$. This recovers the exact exponential
dependence $\omega_{\m}(r=0) =\mbox{exp}(-\tfrac{\pi}{8}\tilde{U})$ for the
$r=0$ metallic AIM \cite{b.2}, in turn showing the $1/r$ exponent
of $\omega_{\m}(r)$ to be exact as $r \rightarrow 0$.

 Finally, we require the $\tilde{\omega}$-dependence of
$\tilde{\Sigma}_{\uparrow}^{\subr}(\omega) \equiv \Sigma_{\uparrow}^{\subr}(\omega)
- \Sigma_{\uparrow}^{\subr}(0)$, given using Eq.(6) by

\begin{equation}
(\Delta_{0}\omega_{\m}^{r})^{-1}\tilde{\Sigma}_{\uparrow}^{\subr}(\omega)
= -\gamma (r) [ |1+\tilde{\omega}|^{r} -1 ]
\end{equation}
which as required scales solely in terms of $\tilde{\omega}$. Eqs.(5,7)
for $\tilde{\Sigma}_{\uparrow}(\omega)$ ($=-\tilde{\Sigma}_{\downarrow}
(-\omega)$), together with Eq.(4), determine the entire LMA scaling spectrum;
given for $\tilde{\omega} \geq0$ (${\cal{F}}(\omega) = {\cal{F}}(-\omega)$)
by

\begin{equation}
\begin{split}
{\cal{F}}(\omega)cos^{2}[\tfrac{\pi}{2}r] = &
\frac{\tfrac{1}{2}\tilde{\omega}^{2r}}
{[\beta (r) \tilde{\omega}^{r} + \gamma (r) (|\tilde{\omega}+1|^{r} - 1)]^{2}
+ \tilde{\omega}^{2r}} \quad + \\
&\frac{\tfrac{1}{2}\tilde{\omega}^{2r}[1+
4\theta(\tilde{\omega}-1)|1-\tfrac{1}
{\tilde{\omega}}|^{r}]}
{[\beta (r) \tilde{\omega}^{r} - \gamma (r) (|\tilde{\omega} -1|^{r}-1)]^{2}
+ \tilde{\omega}^{2r}[1+4\theta(\tilde{\omega}-1)
|1-\tfrac{1}{\tilde{\omega}}|^{r}]^{2}}
\end{split}
\end{equation}
with $\beta (r) = \mbox{tan}[\tfrac{\pi}{2}r] \sim \tfrac{\pi}{2}r$ and
$\gamma (r) \sim \tfrac{4}{\pi r}$.
Eq.(8) reproduces the scaling
spectrum obtained numerically in \cite{b.8}, fully quantitatively as $r 
\rightarrow 0$. For $r=0, 0.05, 0.1$ and $0.15$ it is illustrated in Fig.1 on 
\emph{all} relevant $\tilde{\omega}$-scales, now discussed. 
\begin{figure}
\begin{center}
\psfrag{xaxis}[bc][bc]{\large \bf $\tilde{\w}$}
\psfrag{yaxis}[bc][bc]{\large \bf $\scrF(\w)$}
\epsfig{file=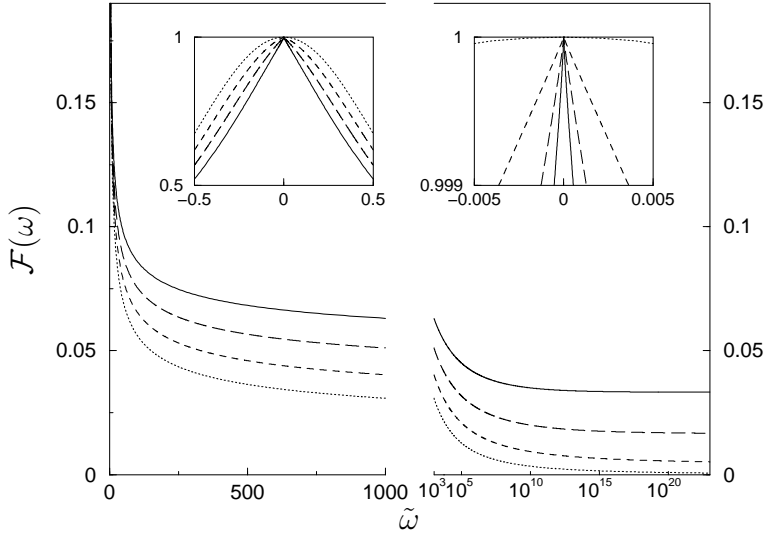,width=10cm} 
\caption{LMA scaling spectrum for the GFL phase (Eq.(8)): $\scrF(\w)$ vs. $\tilde{\w}=\w/\wm$ ($\wm\equiv\w_{\K}$) on all relevant energy scales; for $r=0$ (dotted), 0.05 (dashed), 0.1 (long-dashed) and 0.15 (solid).}
\end{center}
\end{figure}
The leading $\tilde{\omega} \rightarrow 0$ behaviour is ${\cal{F}}(\omega) \sim
1 - 4r\tilde{\omega}^{1-r} - (4/\pi)^{2}\tilde{\omega}^{2(1-r)}$,
with an $|\omega|^{1-r}$ cusp (Fig.1 inset) whose weight vanishes as 
$r \rightarrow 0$; recovering for $r=0$ the conventional Fermi liquid form
for the metallic AIM,  ${\cal{F}}(\omega) - 1 \sim \tilde{\omega}^{2}$.

We turn now to the spectral `tails',
arising formally for $|\tilde{\omega}| \gg 1$ \cite{b.16}.
Based on numerical
fitting of LMA and NRG results it was argued in \cite{b.9} that for all 
$0 \leq r < \frac{1}{2}$, as hitherto believed for $r=0$ (\cite{b.15} and references 
therein), the power-law
decay $\omega_{\m}^{r}D(\omega) \sim |\tilde{\omega}|^{-1/2}$ arose. For the
metallic AIM however, this is now known to be wrong: recent 
work \cite{b.15} shows the tail decay to be far slower, in fact
logarithmically so for all $|\tilde{\omega}| \gtrsim 1$. And the situation
is more subtle for $r \neq 0$. For $|\tilde{\omega}| \gg 1$, but for
$|\tilde{\omega}|^{r} = \mbox{e}^{r\mbox{\ssz{ln}}|\tilde{\omega}|}$ otherwise arbitrary,
Eq.(8) gives

\begin{equation}
{\cal{F}}(\omega) \sim \frac{1}{2} 
\left\{ \frac{1}{[\tfrac{4}{\pi r}(1- \mbox{e}^{-r\mbox{\ssz{ln}}|\tilde
{\omega}|})]^{2} +1}     +
\frac{5}{[\tfrac{4}{\pi r}(1-\mbox{e}^{-r\mbox{\ssz{ln}}|\tilde{\omega}|})]^{2} + 25}  \right \}
\end{equation}
(where the leading low-$r$ behaviour is used). For $r=0$, where
$\tfrac{1}{r} (1-\mbox{e}^{-r\mbox{\ssz{ln}}|\tilde{\omega}|}) = 
\mbox{ln}|\tilde{\omega}|$, this reduces to the known result (Eq.(4.2) of \cite{b.15})
 which in practice 
agrees quantitatively with NRG calculations for $\tilde{\omega} \gtrsim 5$ \cite{b.15}. 
That same form clearly obtains for $r \neq 0$ provided 
$1 \ll |\tilde{\omega}| \ll \mbox{e}^{1/r}$; and for $r \ll 1$ `log-tails' thus
appear to  dominate the spectrum, see Fig.1. But for $r \neq 0$ the ultimate 
asymptote of ${\cal{F}}(\omega)$, approached for $|\tilde{\omega}|^{r}
\gg 1$ (see Fig.1), is a non-zero \emph{constant}:
${\cal{F}}(\omega) \sim (3 \pi^{2}/16) r^{2} 
= (3 \pi^{2}/16) [\rho_{0}J_{\cc}]^{2}$; and hence

\begin{equation}
\pi \Delta_{0} \omega_{\m}^{r} D(\omega) \sim
\tfrac{3 \pi^{2}}{16} r^{2} |\tilde{\omega}|^{-r}
\end{equation}
--- where, since $\tilde{\omega} = \omega/\omega_{\m}$, we note that
the Kondo scale $\omega_{\m}$ `drops out' of the tail. Corresponding
results for the single self-energy $\Sigma(\omega)$ are also readily obtained
via $\Sigma(\omega) = \omega - \Delta(\omega) - 1/G(\omega)$;
such that $(\Delta_{0}\omega_{\m}^{r})^{-1}\Sigma(\omega)$ scales universally in
$\tilde{\omega}$. For $1 \ll |\tilde{\omega}| \ll \mbox{e}^{1/r}$ one finds the
intermediate behaviour $(\Delta_{0}\omega_{\m}^{r})^{-1}\Sigma^{\subi}(\omega) \sim
(16/3 \pi^{2})\mbox{ln}^{2}(|\tilde{\omega}|)$, just as for $r=0$ \cite{b.15}; but its
ultimate asymptotic form for $|\tilde{\omega}|^{r} \gg 1$ is

\begin{equation}
(\Delta_{0}\omega_{\m}^{r})^{-1}\Sigma^{\subi}(\omega) \sim
(16/3 \pi^{2}r^{2}) |\tilde{\omega}|^{r}
\end{equation}
(with $\Sigma^{\subr}(\omega) \sim
-\mbox{sgn}(\omega)\beta (r)\Sigma^{\subi}(\omega)$),
from which $\omega_{\m}$ again drops out.

\section{LM scaling spectrum} But what of the LM phase $\tilde{U} >
\tilde{U}_{\cc}(r)$, throughout which (as above, see also \cite{b.8}) the Kondo spin-flip
scale $\omega_{\m} =0$; and where for $|\omega| \rightarrow 0$, $D(\omega)$
is known to \emph{vanish} as $D(\omega) \propto |\omega|^{r}$ \cite{b.6,b.8,b.9}? 
Here, since $\omega_{\m} =0$,
$\Sigma_{\uparrow}(\omega) = U^{2} {\cal{G}}_{\downarrow}^{-}(\omega)$
in strong coupling ($\tilde{U} \gg 1$), and hence
$\Delta_{0}^{-1}\Sigma_{\uparrow}^{\subi}(\omega) = \theta(-\omega)4|\omega|^{r}$
(\emph{i.e.} Eq.(5) with $\omega_{\m} =0$);  likewise
$\Delta_{0}^{-1}[\tilde{\Sigma}_{\uparrow}^{\subr}(\omega) -
\tilde{\Sigma}_{\uparrow}^{\subr}(0)] = -\gamma(r)|\omega|^{r}$ (Eq.(7) with
$\omega_{\m} =0$). But in the LM phase symmetry is not restored, hence
the `renormalized level' $\tilde{\Sigma}_{\uparrow}^{\subr}(0) = -\tfrac{U}{2}
+ U^{2}$Re${\cal{G}}_{\downarrow}^{-}(0) \neq 0$; and is given via Eq.(6)
by $\Delta_{0}^{-1}\tilde{\Sigma}_{\uparrow}^{\subr}(0) = -\tfrac{1}{2}
[\tilde{U} - \tilde{U}_{\cc}]$, vanishing linearly as $\tilde{U}
\rightarrow \tilde{U}_{\cc}+$. It is now simple to show that
the LM phase is itself characterized by a non-zero low-energy scale 
$\omega_{\mbox{\ssz{l}}} \equiv \omega_{\mbox{\ssz{l}}}(r)$ that is determined by the renormalized level and
given by 
$\omega_{\mbox{\ssz{l}}} = [(\Delta_{0}\gamma(r))^{-1}|\tilde{\Sigma}_{\uparrow}^{\subr}
(0)|]^{1/r}$; whence 
$\omega_{\mbox{\ssz{l}}}= ([\tilde{U}/\tilde{U}_{\cc}] -1)^{1/r}$ using
$\tilde{U}_{\cc}(r)/2\gamma(r) = 1$ for small $r$ (which form, for $\frac{1}{2}<r\leq 1$, has also been found in a recent NRG study \cite{b.11} close to the asymmetric QCP of the model).  Close to the (symmetric) QCP we find that $\omega_{\mbox{\ssz{l}}}^{r}D(\omega)$ indeed scales universally in terms of
$\tilde{\omega} = \omega/\omega_{\mbox{\ssz{l}}}$, and is given explicitly by:
 
\begin{equation}
\pi\Delta_{0}\omega_{\mbox{\ssz{l}}}^{r}D(\omega) =  \tfrac{3 \pi^{2}}{16} r^{2}
\frac{|\tilde{\omega}|^{r}}{(1+|\tilde{\omega}|^{r})^{2}}
\end{equation} 
The LM and GFL phases are thus \emph{each}
characterized by a low-energy scale $\omega_{*} = |[\tilde{U}/\tilde{U}_{\cc}]
-1|^{1/r}$ for small $r$, reducing to $\omega_{\m}$ [$\omega_{\mbox{\ssz{l}}}$] for
$\tilde{U} < \tilde{U}_{\cc}$ [$>\tilde{U}_{\cc}$] in the GFL [LM] phase; and
vanishing as the QCP is approached from either phase. Eq.(12) recovers 
the known lowest-$\omega$ behaviour \cite{b.8}, $\pi\Delta_{0}D(\omega) \sim
3(\Delta_{0}/\tilde{\Sigma}_{\uparrow}^{\subr}(0))^{2}|\omega|^{r}$.
And for $|\tilde{\omega}|^{r} \gg 1$, noting that
$\omega_{\m} \equiv \omega_{*} \equiv \omega_{\mbox{\ssz{l}}}$ as above, Eq.(12)
for the LM phase reduces precisely to Eq.(10) for the GFL phase: the
scaling spectra of the two phases share common tails.

\section{Quantum critical behaviour} Precisely at the QCP, the low-energy 
scale $\omega_{*} =0$. In consequence the low-$\omega$ behaviour of
$D(\omega)$ is given by

\begin{equation}
\pi\Delta_{0}D(\omega; U=U_{\cc}(r)) = \tfrac{3 \pi^{2}}{16}r^{2}|\omega|^{-r}
= \tfrac{3 \pi^{2}}{16} (\rho_{0}J_{\cc})^{2}|\omega|^{-r}
\end{equation}
as follows by taking the limit $\omega_{*} =  0$ in either
Eq. (10) or (12); or equivalently by employing $\omega_{*}=0$ from the outset.
The important point is that at the QCP \emph{Eq.(13) holds as $|\omega|
\rightarrow 0$}, with a prefactor to the $|\omega|^{-r}$ divergence that is 
{\it renormalized} from its non-interacting value (of $\mbox{cos}^{2}[\tfrac{\pi}{2}r]
\sim 1$) to precisely the high-energy 
plateau value of ${\cal{F}}(\omega)$ in the GFL phase, found above.
Physically this embodies the fact that as 
$\tilde{U} \rightarrow \tilde{U}_{\cc}(r)$, the Kondo resonance in
${\cal{F}}(\omega)$ collapses at the QCP, leaving only the tails and
producing a featureless ${\cal{F}}(\omega)$ that is no longer pinned to unity
at the Fermi level, ${\cal{F}}(0) = (3\pi^{2}/16)r^{2}$. In this way
${\cal{F}}(\omega)$ (and hence $D(\omega)$) \emph{at} the QCP is distinct
from that \emph{approaching} it, where $\omega_{\m} \equiv \omega_{*}$
is small but strictly non-zero and any rescaling in terms of it recovers Eq.(8)
(and hence ${\cal{F}}(0)=1$).
The non-[G]FL, interacting nature of the fixed point \cite{b.12} is
clearly evident; embodied both in ${\cal{F}}(0) \neq 1$ and, relatedly,
in the $|\omega| \rightarrow 0$ behaviour 
$(\Delta_{0})^{-1}\Sigma^{\subi}(\omega) = (16/3\pi^{2}r^{2})|\omega|^{r}$
(Eq.(11) with $\omega_{\m}=0$), whose decay to zero with the same power
as the hybridization vitiates the necessary condition \cite{b.8,b.10} for
adiabatic continuity to the non-interacting limit and hence  
a GFL state.

\begin{figure}
\begin{center}
\epsfig{file=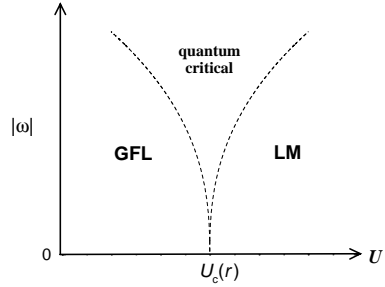,width=5.0cm} 
\caption{Schematic in the $(|\w|,U)$ plane of spectral behaviour in the vicinity of the QCP, with $\w=0$ the Fermi level.  The crossover scale $\w_*$ (see text) is shown as  a dashed line.}
\end{center}
\end{figure}

In summary we have considered a local moment approach to the pseudogap Anderson model close to the symmetric quantum critical point; focusing on $T=0$ single-particle dynamics, in which the collapse of the Kondo resonance as the QCP is approached is directly manifest (the present approach does not enable us to handle on a comparable footing the local magnetic susceptibility $\chi''_{\mbox{\ssz{loc}}}$ considered in \cite{b.12}).  Local impurity spectra close to the QCP have been shown to exhibit scaling in terms of a low-energy scale,
$\omega_{*} = |[\tilde{U}/\tilde{U}_{\cc}]-1|^{1/r}$ for small $r$,
that vanishes as the QCP is approached from either phase;
 and scaling spectra in both phases have been obtained analytically.  The situation is summarised in Fig. 2, representing a natural complement to the conventional picture in the ($T$,$U$)-plane that schematises the crossover behaviour of static properties in the vicinity of a QCP \cite{b.13a}.  Here by contrast, crossovers in dynamics are exemplified in the ($|\w|$,$U$)-plane at $T=0$: for $|\w|\ll \w_{*}$ the low-energy spectral characteristics of the GFL and LM phases are distinct, while for $|\w|\gg \w_{*}$ each reduces to the {\it common} form $\pi\Delta_0D(\w)\sim (3\pi/16)(\rho_0J_{\cc})^2|\w|^{-r}$; such that the latter behaviour, precisely that of the QCP itself, ever increasingly dominates the dynamics of either phase as the transition is approached and $\w_{*}$ vanishes.  The pseudogap AIM is itself a paradigm for understanding generic aspects of local quantum phase transitions, see e.g.\ [11-13]; and we are currently extending our approach to encompass the asymmetric case and the influence of a local magnetic field.

\acknowledgments
We thank K. Ingersent for helpful discussions, and the EPSRC and
Leverhulme Trust for financial support.


\begin{thebibliography}{10}
\bibitem{b.16}
 {As for $r=0$ \cite{b.15}, our results are not specific to use of the p-h ladder sum for $\Pi^{+-}(\omega)$}.
\end{thebibliography}
\end{document}